\documentclass[12pt]{iopart}

\usepackage{bm}
\usepackage{color}
\usepackage{epsfig}
\usepackage{citesort}
\usepackage{graphicx}
\eqnobysec

\newcommand{\lwig}{\mbox{\;\raisebox{.3ex}
    {$<$}$\!\!\!\!\!$\raisebox{-.9ex}{$\sim$}\;}}

\newcommand{\lambdabar}{{\hbox{$\lambda$\kern-1.ex\raise+0.45ex\hbox{--}}}}
\newcommand{\Msun}{M_{\odot}}

\DeclareMathAlphabet{\mathpzc}{OT1}{pzc}{m}{it}

\begin{document}

\begin{flushright}
{\large \tt
TTK-10-47}
\end{flushright}

\title{Spherical collapse of dark energy with an arbitrary sound speed}

\author{Tobias Basse$^1$, Ole Eggers Bj{\ae}lde$^{1,2}$ and Yvonne Y.~Y.~Wong$^2$}
\address{$^1$ Department of Physics and Astronomy, Aarhus University,
Ny Munkegade 120, DK-8000 Aarhus C, Denmark}
\address{$^2$ Institut f\"ur Theoretische Teilchenphysik und Kosmologie \\ RWTH Aachen,
D-52056 Aachen, Germany} \ead{\mailto{tb06@phys.au.dk,
oeb@phys.au.dk, yvonne.wong@physik.rwth-aachen.de}}

\begin{abstract}
We consider a generic type of dark energy fluid, characterised by a
constant equation of state parameter $w$ and sound speed $c_s$, and
investigate the impact of dark energy clustering on cosmic structure
formation using the spherical collapse model.
Along the way, we also discuss in detail the evolution of dark energy perturbations
in the linear regime.
We find that the
introduction of a finite sound speed into the picture necessarily
induces a scale-dependence in the dark energy clustering, which in
turn affects the dynamics of the spherical collapse in a
scale-dependent way. As with other, more conventional fluids, we can
define a Jeans scale for the dark energy clustering, and hence a
Jeans mass $M_J$ for the dark matter which feels the effect of dark
energy clustering via gravitational interactions.  For bound objects
(halos) with masses $M \gg M_J$, the effect of dark energy
clustering is maximal.  For those with $M \ll M_J$, the dark energy
component is effectively homogeneous, and its role in the formation
of these structures is reduced to its effects on the Hubble
expansion rate.  
To compute quantitatively the virial density and the linearly extrapolated threshold density, 
we use a quasi-linear approach which is expected to be valid up to around the Jeans mass.
We find an interesting dependence of these quantities 
on the halo mass $M$, given some $w$ and $c_s$.  The dependence is
the strongest for masses lying in the vicinity of $M \sim M_J$.
Observing this $M$-dependence will be a tell-tale sign that dark
energy is dynamic, and a great leap towards pinning down its
clustering properties.

\end{abstract}

\maketitle

\section{Introduction}

The apparent accelerating expansion of our universe is, according to
the standard model of cosmology, best described by the presence of a
dark energy component with a strong negative pressure, dominating
the gravitational physics on large scales (see, e.g.,
\cite{Copeland:2006wr,Frieman:2008zz,Bean:2010xw} for reviews).
However, our knowledge of the actual properties of this dark energy
is very limited, and a number of open questions remain---Is dark
energy dynamic or not? Does it take part in clustering? Is the dark
energy's behaviour scale-dependent?---to name but a few.

In order to distinguish between various models of dark energy, we
must appeal to a variety of observational tests, each probing a
different aspect of the dark energy's dynamics.  Luminosity and
angular diameter distance measurements using, respectively, type Ia
supernovae and the baryon acoustic oscillation scale have
yielded---and will continue to yield---interesting information on
the dark energy's influence on the expansion history of the
universe. The cosmic microwave background temperature
anisotropies~\cite{Weller:2003hw,DeDeo:2003te,Bean:2003fb,Hannestad:2005ak,dePutter:2010vy},
as well as their cross-correlations with tracers of the large-scale
structure
distribution~\cite{Hu:2004yd,Corasaniti:2005pq,Giannantonio:2008zi,Li:2010ac},
provide a means to track the dark energy's impact on the evolution
of the gravitational potential via the  late integrated Sachs--Wolfe
effect.
Weak gravitational lensing of distant objects probes the dark
energy's effect on the distance--redshift relation and the growth
function~\cite{Weinberg:2002rd,Jarvis:2005ck,Schimd:2006pa}.
Lastly,  the formation of gravitationally bound objects such as
galaxies and galaxy clusters is also sensitive to the detailed
properties of the dark energy
component~\cite{Takada:2006xs,Sefusatti:2006eu,Abramo:2009ne,Alimi:2009zk}.

Into the last category falls the so-called spherical collapse
model~\cite{Gunn:1972sv}, which, as the name suggests, is a model of
gravitational collapse simplified by the assumption of spherical
symmetry.  In the model, a spherically symmetric overdense region
with uniform density evolves to a configuration of infinite density
under its own gravity, and  a gravitationally bound object is said
to be formed.

The original spherical collapse model was constructed under the
assumption of an Einstein--de Sitter (EdS) universe. Already by the
1980s and the early 1990s the model had been extended to include a
cosmological
constant~\cite{Peebles:1984ge,Weinberg:1987dv,Lahav:1991wc},  and
later for quintessence~\cite{Wang:1998gt}.  More analyses have
followed since then, and all reached the conclusion that dark energy
has an important impact on the formation of gravitationally bound
structures~\cite{Mota:2004pa,Horellou:2005qc,Bartelmann:2005fc,Abramo:2007iu}.
The topic appears to have gained  momentum again during the past
year~\cite{Basilakos:2009mz,Lee:2009qq,Creminelli:2009mu,Pace:2010sn,Wintergerst:2010ui,Valkenburg:2011tm}
mainly because of the positive expectation that these results will
be testable against observations in the not-too-distant
future~\cite{Albrecht:2006um}.

In this paper we investigate by means of the spherical collapse
model how a generic dark energy component characterised by a
constant equation of state parameter $w< -1/3$ and sound speed $c_s$
affects the formation of cosmic structures. In previous studies the
dark energy component is generally allowed to take on different
values of $w$.  However, the sound speed $c_s$ is invariably assumed
to be either approaching the speed of light so that the dark energy
is essentially homogeneous, or exactly vanishing so that the dark
energy fluid is comoving with the matter component.  These
assumptions undoubtedly simplify the calculations considerably and
represent the two limiting behaviours of dark energy clustering.
However, the case of a  general sound speed remains interesting in
that it introduces a scale-dependence into the problem.  Identifying
this dependence will give us yet another pointer to the true nature
of dark energy.  Here, we demonstrate for the first time how to
incorporate a dark energy component with an arbitrary sound speed
into the spherical collapse model (in an approximate way).

The paper is organised as follows. In section~\ref{sec:II} we
introduce the evolution equations for the spherical collapse and the
corresponding equations of motion for the dark energy component.  In
section~\ref{sec:linear} we discuss dark energy clustering within
the framework of linear perturbation theory.  Section~\ref{sec:III}
contains our numerical results.  Our conclusions are presented in
section~\ref{sec:IV}.

\section{Spherical collapse\label{sec:II}}

\subsection{The spherical top hat and equations of motion for the matter component}

In its most basic formulation, the spherical collapse model assumes
there exists a spherically symmetric overdense region on top of an
otherwise uniform  background matter density field. The overdense
region is characterised by a physical radius $R_i \equiv R(\tau_i)$
at the initial (conformal) time $\tau_i$, and a uniform initial
energy  density
\begin{equation}
\rho^{\rm th}_m(\tau_i)  \equiv \bar{\rho}_m(\tau_i)(1+\delta^{\rm th}_{m,i}),
\end{equation}
where $\bar{\rho}_m(\tau)$ denotes the energy density of the
background matter field.  This is our spherical ``top hat''
perturbation, and the mass contained within is given by
\begin{equation}
\label{eq:sphericalm}
M = \frac{4 \pi}{3} {\bar \rho}_m (\tau_i) (1 + \delta^{\rm th}_{m,i}) R_i^3
=\frac{4\pi}{3} \bar{\rho}_m(\tau_0) (1 + \delta^{\rm th}_{m,i}) X_i^3,
\end{equation}
where $\tau_0$ denotes the present time, and we have defined
\begin{equation}
X \equiv \frac{R}{a}
\end{equation}
as the comoving radius of the top hat.

The evolution of the physical top hat radius $R$ with respect to
{\it cosmic time} $t$ is described by the familiar equation of
motion
\begin{equation}
\label{eq:ddotr}
\frac{1}{R}\frac{d^2 R}{dt^2} = - \frac{4 \pi G}{3} (\rho^{\rm th}_m + \rho^{\rm th}_Q +3 P_Q^{\rm th}),
\end{equation}
where we have incorporated in the equation the possibility of a
second energy component with a nonzero pressure denoted by the
subscript $Q$.  This second component is uniform inside the top hat
region defined by the radius $R$, and shall be our dark energy
component in this work.  Equation~(\ref{eq:ddotr}) can be
equivalently expressed as an equation of motion for for the comoving
top hat radius  $X$ with respect to conformal time $\tau$,
\begin{equation}
\frac{\ddot X}{X} + {\cal H} \frac{\dot X}{X} = - \frac{4 \pi G}{3}a^2 [\bar{\rho}_m \delta^{\rm th}_m  + \bar{\rho}_Q (1+3 c_s^2) \delta^{\rm th}_Q],
\label{eq:xdotdot}
\end{equation}
where $\cdot \equiv \partial/\partial \tau$,  ${\cal H}=a H$ is the
conformal Hubble parameter, and
\begin{equation}
\label{eq:cs2}
c_s^2 \equiv \left.  \frac{\delta P_Q}{\delta \rho_Q} \right|_{\rm rest}
\end{equation}
is the square of dark energy sound speed defined in the dark energy
fluid's rest frame. Note that in identifying $c_s$ in
equations~(\ref{eq:xdotdot}) and~(\ref{eq:cs2}) as the rest frame
sound speed, we have implicitly assumed that we are dealing only
with length scales much smaller than the Hubble length, and that averaged over the spherical region, 
there is no bulk flow of dark energy relative to the dark matter fluid.
We also assume $c_s^2$ to
be constant in time and space.

Since the total mass of matter inside the top hat  $M= (4 \pi/3)
\rho_m^{\rm th} R^3$ is conserved, the top hat matter density
contrast $\delta^{\rm th}_m$ can be easily  expressed as a
function of the top hat radius,
\begin{eqnarray}
\label{eq:deltamth}
\delta^{\rm th}_m (\tau) & \equiv & \frac{\rho_m^{\rm th}(\tau)}{\bar{\rho}_m(\tau)}-1 \nonumber \\
&=& (1 + \delta^{\rm th}_{m,i}) \!
\left[\frac{a(\tau)}{a(\tau_i)} \frac{R_i}{R(\tau)} \right]^3  -1= (1 + \delta^{\rm th}_{m,i})  \!
\left[\frac{X_i}{X(\tau)} \right]^3  -1.
\end{eqnarray}
For the dark energy density contrast $\delta_Q^{\rm th}$, two
limiting cases have been studied in the literature. The first is the
non-clustering limit, in which the dark energy sound speed $c_s$ is taken
to approach the speed of light, see e.g. \cite{Wang:1998gt}. In this
case, $\delta^{\rm th}_Q$ is effectively zero, so that the role of
dark energy in the spherical collapse enters only through the Hubble
expansion of the background.%
\footnote{Note that in order to get
exactly zero dark energy clustering, the sound speed would have to
be infinite; See the linear solutions~(\ref{eq:clustering}) and~(\ref{eq:nonclustering}).}

The second is the ``comoving'' or clustering limit, in which the
dark energy sound speed is exactly zero, see e.g.
\cite{Creminelli:2009mu}. As we shall see in the next section,  the
Euler equation for the dark energy fluid in this limit is identical
to its counterpart for a nonrelativistic  dark matter fluid.  This
means that the bulk velocity fields of the two fluids are the same;
the fluids are thus said to be comoving.  Note that this observation
does not imply the dark energy and dark matter density contrasts
evolve in the same manner, since the conditions for energy
conservation differ between the two fluids.   It does, however,
imply a conservation law  for the dark energy component inside the
top hat, so that the evolution of  $\rho^{\rm th}_Q$ can be simply
expressed as
\begin{equation}
\frac{d \rho^{\rm th}_Q}{d t} + \frac{3}{R} \frac{dR}{dt} ( \rho_Q^{\rm th} + \bar{P}_Q) =0,
\end{equation}
or
\begin{equation}
\dot{\rho}^{\rm th}_Q +3 \left( {\cal H} + \frac{\dot{X}}{X} \right) ( \rho_Q^{\rm th} + \bar{P}_Q) =0
\end{equation}
in terms of comoving quantities.

Strictly speaking, these two limiting cases are the only ones for
which the top hat formulation is exact. The case of an arbitrary
dark energy sound speed $c_s$ is strictly not amenable to this
simple treatment, since the existence of a finite sound speed and
therefore the provision for the propagation of sound waves imply
that the energy densities---both dark matter and dark
energy---inside the overdense region must evolve to a nonuniform configuration,
even if they are initially uniform.  Having
said this, however, we must also bear in mind that the spherical
collapse model is itself a simplified model of structure formation,
and the top hat density contrast should be interpreted as the
average density contrast inside a region after a unit top hat
filtering function has been applied.
If we take this as our guiding principle, then the generalisation of
the spherical top hat to include a dark energy component  with an
arbitrary sound speed simply requires that we interpret $\delta^{\rm
th}_Q$ as the spatially averaged density contrast of the dark energy
field inside a region of comoving radius $X$.
 Symbolically, this spatial average can be expressed as
\begin{equation}
\label{eq:detophat}
\delta^{\rm th}_Q (\tau) \equiv  \frac{3}{X^3} \int_0^{X} dx \ x^2  \delta_Q(x,\tau),
\end{equation}
where $x \equiv |{\bm x}|$, and ${\bm x}$  denotes the comoving coordinates.

\subsection{Equations of motion for the dark energy component\label{sec:eom}}

It remains to specify an evolution equation for the dark energy
density perturbation $\delta_Q(x,\tau)$. We begin by writing down
the continuity and Euler equations for a relativistic fluid $\alpha$
in an expanding background in the pseudo-Newtonian approach~\cite{Peeblesbook},
\begin{eqnarray}
\label{eq:eom}
&&{\dot \rho}_\alpha + 3 {\cal H} (\rho_\alpha + P_\alpha) +  \nabla \cdot [(\rho_\alpha+P_\alpha) {\bm u}_\alpha] =0, \nonumber \\
&& {\dot {\bm u}}_\alpha + {\cal H} {\bm u}_\alpha +  ({\bm u}_\alpha \cdot \nabla) {\bm u}_\alpha +\frac{\nabla P_\alpha+{\bm u}_\alpha {\dot P}_\alpha}{\rho_\alpha+P_\alpha} +
\nabla \phi_N=0.
\end{eqnarray}
Here,   $\nabla \equiv \partial/\partial {\bm x}$, ${\bm u}_\alpha$
is the peculiar velocity of the fluid,  and the potential $\phi_N$ can
be obtained from the Poisson equation
\begin{equation}
\label{eq:poisson}
\nabla^2 \phi = 4 \pi G a^2 \sum_\alpha \delta \rho_\alpha + 3 \delta P_\alpha.
\end{equation}
These equations should apply if we restrict our considerations to (i) length scales
much smaller than the Hubble length, (ii) nonrelativistic peculiar velocities, and (iii) 
nonrelativistic sound speeds $c_s \ll 1$.
We demonstrate in \ref{sec:rel} that, at the linear level, these equations are indeed consistent with the Newtonian limit of 
a  general relativistic formulation (see, e.g.,~\cite{Sapone:2009mb,Ballesteros:2010ks,Anselmi:2011ef}).

Defining the equation of state parameter for the dark energy component
\begin{equation}
w \equiv \frac{\bar{P}_Q}{\bar{\rho}_Q},
\end{equation}
and using the definitions $P_\alpha \equiv \bar{P}_\alpha + \delta P_\alpha$ and  $\delta_\alpha^P \equiv \bar{\rho}_\alpha^{-1} \delta P_\alpha$,
equation~(\ref{eq:eom}) can be rewritten for $\alpha=Q$ as
\begin{eqnarray}
&&\dot{\delta}_Q + 3 {\cal H} (\delta_Q^P - w \delta_Q) 
+
\nabla \cdot [(\rho_Q+P_Q) {\bm u}_Q/\bar{\rho}_Q]=0, \nonumber \\
&& \dot{\bm u}_Q + {\cal H} {\bm u}_Q + ({\bm u}_Q \cdot \nabla) {\bm u}_Q +\frac{ \nabla \delta_Q^P+ {\bm u}_Q (w \dot{\bar{\rho}}_Q/\bar{\rho}_Q+ \dot{\delta}_Q^P)}{1+w+\delta_Q +  \delta_Q^P}+
\nabla \phi_N=0. 
\label{eq:deltaqfourier}
\end{eqnarray}
 Since in our set-up the universe contains
only dark matter and dark energy, the Poisson
equation~(\ref{eq:poisson}) now reads
\begin{equation}
\label{eq:poisson2}
\nabla^2 \phi = 4 \pi G a^2 [ \bar{\rho}_m \delta_m + \bar{\rho}_Q  (\delta_Q+ 3 \delta_Q^P) ], 
\end{equation}
where, for our particular problem, the dark matter density
perturbation $\delta_m(x,\tau)$ takes the form
\begin{equation}
\label{eq:deltam}
\delta_m(x,\tau) = \left\{ \begin{array}{ll}
                \delta_m^{\rm th}(\tau), & x \leq X(\tau), \\
                0, &    x > X(\tau). \end{array} \right.
\end{equation}
with the top hat density contrast $\delta_m^{\rm th}(\tau)$ given by
equation~(\ref{eq:deltamth}).

The continuity and Euler equations~(\ref{eq:deltaqfourier}) are
nonlinear in the quantities $\delta_Q$ and ${\bm u}_Q$,
 which even under the assumption of  spherical symmetry are nontrivial to solve.
Therefore, as a first approximation, we linearise them to obtain
\begin{eqnarray}
\label{eq:linearq}
&&\dot{\delta}^{\rm lin}_Q + 3 {\cal H} (\delta_Q^{P,{\rm lin}} - w \delta^{\rm lin}_Q)  + (1+w)
\theta^{\rm lin}_Q=0, \nonumber \\
&& \dot{\theta}^{\rm lin}_Q + (1-3 w) {\cal H} \theta^{\rm lin}_Q  +\frac{\nabla^2 \delta^{P,{\rm lin}}_Q}{1+w} +
\nabla^2 \phi_N=0,
\end{eqnarray}
where we have defined the divergence of the dark energy velocity
field to be $\theta_Q \equiv \nabla \cdot {\bm u}_Q$, and assumed $w$  to be constant in time.
 Linearisation
assumes that the perturbed quantities $\delta_Q$ and $\theta_Q$ are
small.  This is likely a good assumption since (i) the presence of a
finite sound speed $c_s$ naturally hinders the clustering of dark
energy, keeping $\delta_Q \simeq \delta_Q^{\rm lin}$ small relative
to $\delta_m$, and (ii) even in the limit $c_s=0$ where the
clustering of dark energy is most efficient, a fully linear analysis
shows that $\delta_Q^{\rm lin}$ is suppressed relative to
$\delta_m^{\rm lin}$ because of the dark energy's  negative equation
of state parameter (see section~\ref{sec:linear}). Either way, the
assumption of linearity in dark energy clustering can be easily
checked {\it a posteriori} against solutions of the evolution
equations for consistency. Finally, let us stress again that we are
linearising {\it only} the dark energy equations of motion; the
evolution of the dark matter component is still fully nonlinear, and
described by the spherical collapse detailed in the previous
section.  We shall call this the ``quasi-nonlinear'' approach.

Upon linearisation, it is useful to recast the equations of motion
in Fourier space. Define the Fourier transform for some field
$A(x,\tau)$ as
\begin{eqnarray}
A(x,\tau)& =& \frac{1}{(2 \pi)^3} \int d^3 k \ \tilde{A}(k,\tau) \exp(i {\bm k} \cdot {\bm x}) \nonumber \\
&=& \frac{1}{2 \pi^2} \int d k \ k^2 \tilde{A}(k,\tau) \frac{\sin(k x)}{k x}.
\end{eqnarray}
Then, using the Poisson equation~(\ref{eq:poisson2}) and the relation $\tilde{\delta}_Q^{P,{\rm lin}} = c_s^2 \tilde{\delta}_Q^{\rm lin} + 3 {\cal H}(1+w) (c_s^2-w) \tilde{\theta}^{\rm lin}_Q/k^2$~\cite{Bean:2003fb}, equation~(\ref{eq:linearq}) can be equivalently expressed as
\begin{eqnarray}
\label{eq:linearFourier}
&& \dot{\tilde\delta}^{\rm lin}_Q + 3 (c_s^2 - w){\cal H}  \tilde\delta^{\rm lin}_Q 
+ (1+w)
{\tilde\theta}^{\rm lin}_Q=0,   \\
&& \dot{\tilde\theta}^{\rm lin}_Q +(1-3 c_s^2)  {\cal H}
\tilde\theta^{\rm lin}_Q  -\frac{k^2 c_s^2}{1+w} \tilde\delta^{\rm lin}_Q+
 4 \pi G a^2 [ \bar{\rho}_m \tilde\delta_m + \bar{\rho}_Q ( 1 + 3 c_s^2) \tilde\delta^{\rm lin}_Q]
=0,\nonumber 
\end{eqnarray}
where we have dropped subdominant terms proportional to ${\cal H}^2/{k^2}$, since 
we are interested only in subhorizon scales $k \gg {\cal H}$.   For
the dark matter density contrast $\delta_m(x,\tau)$ given in
equation~(\ref{eq:deltam}), the Fourier space equivalent is
\begin{eqnarray}
\tilde\delta_m(k,\tau) & = & 4 \pi \int^{X(t)}_0 dx  \ x^2  \  \delta_m^{\rm th}(\tau)  \frac{\sin(k x)}{k x} \nonumber \\
&=& \delta_m^{\rm th}(\tau) \frac{4 \pi}{3} X^3 W(k X) = \frac{4 \pi}{3} [(1+\delta^{\rm th}_{m,i}) X_i^3 - X^3] W(kX),
\end{eqnarray}
where
\begin{equation}
W(k X) = \frac{3}{(kX)^3} [\sin (kX) - kX \cos (kX)]
\end{equation}
by convention.

Lastly, we would like to relate $\tilde{\delta}^{\rm lin}_Q(k,\tau)$
to the average dark energy density contrast inside the top hat,
$\delta^{\rm th}_Q(\tau)$, as defined in
equation~(\ref{eq:detophat}), since this is the quantity that
ultimately governs the evolution of the top hat radius $X$ via
equation~(\ref{eq:xdotdot}). This step is simple: we only need to
identify $\delta_Q(x, \tau)$ with $\delta_Q^{\rm lin}(x, \tau)$, the
latter of which is obtained by Fourier transforming
$\tilde{\delta}^{\rm lin}_Q(k,\tau)$.  Thus,
equation~(\ref{eq:detophat}) simplifies to
\begin{equation}
\delta_Q^{\rm th}(\tau) = \frac{1}{2 \pi^2} \int dk \ k^2 W(kX) \tilde\delta^{\rm lin}_Q(k,\tau).
\label{eq:deltaqreal}
\end{equation}
Our set of equations of motion is now complete.

\section{Linear theory\label{sec:linear}}

Before we present the results of the spherical collapse model, let
us first consider the evolution of dark matter and dark energy
perturbations in the linear regime, i.e., where the dark matter
perturbations are also tracked with linearised equations of motion.
This exercise is useful for two reasons. Firstly, as we shall see,
an understanding of the linear evolution can shed light on many
essential features of the dependence of dark energy clustering on
its equation of state parameter $w$ and sound speed $c_s$.  At the
same time, the linear solution also sets the initial conditions for
the spherical collapse model.

Secondly, some semi-analytic theories of structure formation such as
the Press--Schechter formalism~\cite{Press:1973iz} and the excursion
set theory~\cite{PH90,Bond:1990iw,Maggiore:2009rv,Maggiore:2009rw}
require as an input a linear critical density contrast $\delta^{\rm
lin}_{\rm coll}$.  In these theories a collapsed structure is
assumed to have formed once the linearly evolved matter density
contrast reaches the threshold value $\delta^{\rm lin}_{\rm coll}$
at some time $\tau_{\rm coll}$. The value of  $\delta^{\rm lin}_{\rm
coll}$ can be determined from the spherical collapse model by
interpreting $\tau_{\rm coll}$ as the instant at which the top hat
radius vanishes. In practice, this means that in order to extract
$\delta^{\rm lin}_{\rm coll}$ for a particular cosmological model,
we need to solve {\it both} the nonlinear and the linear equations
of motion at the same time.

We have already written down the linearised equations of motion for
the dark energy perturbations in equations~(\ref{eq:linearq}) and
(\ref{eq:linearFourier}). For the dark matter component, the
corresponding equations  are
\begin{eqnarray}
&&\dot{\delta}^{\rm lin}_m  +
\theta^{\rm lin}_m=0, \nonumber \\
&& \dot{\theta}^{\rm lin}_m + {\cal H} \theta^{\rm lin}_m
+ 4 \pi G a^2 [ \bar{\rho}_m \delta^{\rm lin}_m + \bar{\rho}_Q ( 1 + 3 c_s^2) \delta^{\rm lin}_Q]
=0.
\label{eq:lindeltam}
\end{eqnarray}
These and equation~(\ref{eq:linearFourier})  are solved
simultaneously to determine the evolution of $\delta^{\rm lin}_Q$
and $\delta^{\rm lin}_m$.

\subsection{Dark energy evolution in the linear regime\label{sec:initial}}

Let us consider first the linear evolution of the dark energy
perturbations.  Here, it is convenient to combine the first order
differential equations~(\ref{eq:linearFourier}) for
$\tilde{\delta}^{\rm lin}_Q$ and $\tilde{\theta}^{\rm lin}_Q$ into
one second order differential equation for $\tilde{\delta}^{\rm
lin}_Q$, and also adopt  a new time variable $s \equiv \ln a$.  Assuming
$w$ and $c_s$ to be constant in time, we find
\begin{equation}
\label{eq:harmonic}
\tilde\delta^{{\rm lin}''}_Q +  {\cal D}(s) \tilde\delta^{{\rm lin}'}_Q
+\left[ \frac{k^2 c_s^2}{{\cal H}^2}{\cal X}(s)-\kappa(s) \right] \tilde\delta^{\rm lin}_Q =  \frac{3}{2}(1+w) \Omega_m(s) \tilde\delta^{\rm lin}_m,
\end{equation}
where $' \equiv \partial/\partial s$,  and
\begin{eqnarray}
\label{eq:dkx}
{\cal D}(s) &\equiv & 1+\frac{{\cal H}'}{\cal H}- 3w, \nonumber \\
\kappa(s) &\equiv& 3w\left(1+\frac{{\cal H}'}{\cal H} \right) +\frac{3}{2}(1+w)  \Omega_Q(s), \nonumber \\
{\cal X}(s) & \equiv & 1+3 \frac{{\cal H}^2}{k^2} \left[ 1 + \frac{{\cal H}'}{\cal H} - 3 (c_s^2 -w)
- \frac{3}{2}(1+w) \Omega_Q(s)\right].
\end{eqnarray}
For the cosmological models considered in this work, ${\cal D}(s)$ and $|\kappa(s)|$ are of order unity, while 
${\cal X}(s)\approx 1$ always holds true because of our assumption of $k\gg {\cal H} $.

Equation~(\ref{eq:harmonic}) describes a damped harmonic oscillator
with a driving force sourced by perturbations in the dark matter
fluid. Exact analytic solutions do not exist  for arbitrary
cosmologies.  However, approximate solutions can be constructed in
certain limits:

\begin{enumerate}
\item {\it Clustering limit}.
This is the limit in which $k^2 c_s^2/{\cal H}^2 \ll |\kappa| \sim
1$.  In this case, all coefficients in the differential equation are
of order unity.  It is therefore necessary to specify the exact time
dependence of ${\cal D}(s)$, $\kappa(s)$ as well as
$\tilde\delta^{\rm lin}_m$ in order to find a solution.  Formally
setting $k=0$, the solution is particularly simple  during the
matter domination epoch, where $\Omega_m(s) \simeq 1$, $\Omega_Q(s)
\ll 1$, ${\cal H}'/{\cal H} \simeq -1/2$, $\tilde\delta^{\rm lin}_m
\propto a$ and $\tilde{\theta}_m^{\rm lin} \simeq - {\cal H}
\tilde{\delta}_m^{\rm lin}$.  At
 $s-s_i \gg 1$ it has the asymptotic form
\begin{eqnarray}
\label{eq:clustering}
\tilde{\delta}^{\rm lin}_Q &\simeq & \frac{1+w}{1- 3w} \tilde{\delta}^{\rm lin}_m, \nonumber \\
\tilde{\theta}^{\rm lin}_Q &\simeq & - \frac{\cal H}{1+w} [3(c_s^2-w)+1]  \tilde{\delta}^{\rm lin}_Q  ,
\end{eqnarray}
where we have obtained the solution for $\tilde{\theta}^{\rm lin}_Q$ from the continuity equation
by first differentiating $\tilde{\delta}^{\rm lin}_Q$ with respect to time.

At first glance, the solution~(\ref{eq:clustering}) for $\tilde{\delta}^{\rm lin}_Q$ appears to be at odds with the solution obtained in, e.g., reference~\cite{Ballesteros:2010ks}
in the same limit (i.e., ${\cal H} \ll k \ll {\cal H}_s$, where ${\cal H}_s \equiv {\cal H}/c_s$ is the inverse of the sound horizon, or the ``Jeans wavenumber'' $k_J$ as we define in equation~(\ref{eq:kjeans}) below).  In particular, the solution 
of~\cite{Ballesteros:2010ks} depends explicitly on the sound speed $c_s^2$, whereas our solution does not.  Part of the discrepancy can be traced to the term ${\cal X}(s)$ defined in equation~(\ref{eq:dkx}).  In our analysis we always approximate this term as ${\cal X}(s) =1$, while  some contributions proportional to ${\cal H}^2/k^2$ have been retained in the analysis of~\cite{Ballesteros:2010ks}.

However, we believe that this discrepancy is of little consequence.  As we demonstrate in~\ref{sec:validity}, the ${\cal H} \ll k \ll {\cal H}_s$ limit is well-defined only for those dark energy sound speeds satisfying $c_s^2 \lwig 10^{-3}$.    Thus, from a numerical point of view, our approximate solution
and that of~\cite{Ballesteros:2010ks}, {\it where they are actually applicable}, are consistent with one another to better than one part in a thousand.

\item {\it Non-clustering limit}.
This limit corresponds to $k^2 c_s^2/{\cal H}^2 \gg |\kappa| \sim
1$, which is also the steady-state limit ($|\tilde{\delta}^{{\rm
lin}''}_Q/\tilde{\delta}^{\rm lin}_Q|, |\tilde{\delta}^{{\rm
lin}'}_Q/\tilde{\delta}^{\rm lin}_Q| \ll 1$).  The solution can be
obtained by formally setting $\tilde{\delta}^{{\rm
lin}''}_Q=\tilde{\delta}^{{\rm lin}'}_Q=\kappa=0$.  Defining the
``Jeans wavenumber''
\begin{equation}
\label{eq:kjeans}
k_J \equiv \frac{\cal H}{c_s},
\end{equation}
the steady-state/non-clustering solution then reads
\begin{eqnarray}
\label{eq:nonclustering}
\tilde{\delta}^{\rm lin}_Q &\simeq& \frac{3}{2} (1+w) \ \Omega_m(s) \left( \frac{k_J}{k} \right)^2 \tilde{\delta}^{\rm lin}_m, \nonumber \\
\tilde{\theta}^{\rm lin}_Q &\simeq & - \frac{3 {\cal H}}{1+w} [c_s^2-w \Omega_m(s)] \tilde{\delta}^{\rm lin}_Q.
\end{eqnarray}
Note that, unlike the clustering solution~(\ref{eq:clustering}), the
non-clustering solution is not restricted to the matter domination
epoch. Furthermore, the derivation of~(\ref{eq:nonclustering}) does
not in fact require the assumption of a linear $\tilde{\delta}_m$,
since  $\tilde{\delta}_m$ enters into the differential
equation~(\ref{eq:harmonic}) only through the gravitational
potential $\phi$, and hence the Poisson equation, which is in any
case linear in  $\tilde{\delta}_m$.  This means that the
steady-state/non-clustering solution~(\ref{eq:nonclustering}) would
have been equally valid had we replaced $\tilde{\delta}_Q^{\rm lin}$
with the Fourier transform of the top hat density contrast
 $\delta^{\rm th}_m$.
We shall make use of this solution again later on in the analysis.

The form of the non-clustering solution is akin to those commonly
found in hot or warm dark matter scenarios, in which $k_J$  is
associated with the free-streaming scale of the problem (see, e.g.,
\cite{RingwaldWong}).  However,  since dark energy has a non-zero
$w$ while free-streaming dark matter does not, an extra prefactor
$(1+w)$ is incurred in the solution~(\ref{eq:nonclustering}).

\item {\it Unstable limit}.
So far we have implicitly assumed  $c_s$  to be a real number.  Let us entertain ourselves for a moment
with the possibility of an imaginary dark energy sound speed.   In the limit  $|k^2 c_s^2/{\cal H}^2| \ll 1$,
the dark energy perturbations are described by the same clustering solution as equation~(\ref{eq:clustering}).
Contrastingly, the $|k^2 c_s^2/{\cal H}^2| \gg 1$ limit is unstable.
Formally setting ${\cal D}=\kappa =\tilde{\delta}^{\rm lin}_m= 0$, equation~(\ref{eq:harmonic})
is solved in the matter domination epoch by $\tilde\delta^{\rm lin}_Q = C_1 I_0(\omega \sqrt{a})+
C_2 K_0(\omega \sqrt{a})$, where
$I_0(x)$ and $K_0(x)$ are the zeroth order modified Bessel functions
of the first and the second kind respectively, and  $\omega \equiv 2 k |c_s|/\sqrt{H_0^2 \Omega_m}$.
For $x \equiv \omega \sqrt{a} \gg 1$, the modified Bessel functions have the asymptotic forms
$I_0(x) \simeq \exp(x)/\sqrt{2 \pi x}$ and $K_0(x) \simeq \sqrt{\pi/2x} \exp(-x)$.  Thus, we find for the
linear dark energy density contrast the asymptotic solution
\begin{equation}
\tilde{\delta}^{\rm lin}_Q \sim a^{-1/4} \exp( \omega \sqrt{a}).
\end{equation}
This exponential growth of  $\tilde{\delta}^{\rm lin}_Q $ in turn
sources the evolution of the dark matter density contrast via the
Poisson equation~(\ref{eq:poisson2}).   Consequently,
$\tilde{\delta}^{\rm lin}_m$ also exhibits a similarly explosive and
strongly scale-dependent growth that at first glance appears to be
in conflict with our current understanding of large-scale structure
formation unless $|c_s|$ is very small. We shall therefore not
pursue the case of an imaginary dark energy sound speed  any further
in the present work.

\end{enumerate}

Given the limiting solutions~(\ref{eq:clustering}) and (\ref{eq:nonclustering}), we can try to interpolate between
the clustering and non-clustering regimes using the following (rough) interpolation formulae:
\begin{eqnarray}
\tilde{\delta}^{\rm lin}_Q &=& \frac{1+w}{1- 3w +(2/3) (k/k_J)^2} \tilde{\delta}^{\rm lin}_m, \nonumber  \\
\tilde{\theta}^{\rm  lin}_Q &= & - \frac{\cal H}{1+w}  \left[3(c_s^2-w)+ \frac{1- 3w}{1- 3w + (2/3)(k/k_J)^2}
\right] \tilde{\delta}^{\rm lin}_Q.
\label{eq:interpolate}
\end{eqnarray}
The maximum error is 30\% at $k \sim k_J$.  These interpolation
formulae are valid during the matter domination regime,
and can be used to set the initial conditions for the dark energy component in the spherical collapse model.%
\footnote{The full equation~(\ref{eq:harmonic}) in fact has an exact analytic solution encompassing all three limits discussed
above in terms of Bessel functions in the matter domination epoch.
However, the complexity of the solution rather obscures the simple physics behind the problem.  We therefore do not quote it here.}

Lastly, let us define a ``Jeans mass'' scale analogous to the Jeans
wavenumber $k_J$ given in~(\ref{eq:kjeans}), i.e., the mass scale at
which we expect the effects of the dark energy sound speed to set
in. The Jeans mass is defined here as
\begin{equation}
M_J(a)  \equiv \frac{4 \pi}{3} \bar{\rho}_m(a) \left( \frac{\lambda_J (a)}{2} \right)^3,
\end{equation}
where $\lambda_J \equiv 2 \pi/k_J$.
Evaluating the expression at $a=1$, we find
\begin{equation}
\label{eq:jeansmass}
M_J = 9.7 \times 10^{23} \ \Omega_{m} \  c_s^3 \ h^{-1} M_\odot.
\end{equation}
For example, given $\Omega_{m}=0.3$ and $h=0.7$, we have $M_J = 1.3 \times 10^{16} M_\odot$
for $c_s^2 = 10^{-5}$, and  $M_J = 4 \times 10^{14} M_\odot$ for $c_s^2 = 10^{-6}$ today.
Note that the mass here refers to the mass of the dark matter component, not the dark energy!

\subsection{Linear threshold density}

Our second motivation for considering  linear theory is the computation of the linear threshold density, defined as
\begin{equation}
\delta^{\rm lin}_{\rm coll}  \equiv \delta^{\rm th,lin}_m(\tau_{\rm coll}),
\end{equation}
where $\delta^{\rm th,lin}_m(\tau)$ is the linearly evolved top hat
matter density, and $\tau_{\rm coll}$ is the instant at which  the
top hat radius goes to zero.   As the name implies,   $\delta^{\rm
th,lin}_m(\tau)$ is the linear version of the quantity $ \delta^{\rm
th}_m(\tau)$ defined in equation~(\ref{eq:deltam}), and is tracked
by the equations of motion~(\ref{eq:lindeltam}) upon the
replacements
\begin{eqnarray}
\delta^{\rm lin}_m(x,\tau) &\to& \delta^{\rm th,lin}_m (\tau), \nonumber \\
 \theta^{\rm lin}_m(x,\tau) &\to& \theta^{\rm th,lin}_m (\tau), \nonumber \\
\delta^{\rm lin}_Q(x,\tau) & \to& \delta^{\rm th,lin}_Q(\tau),
\end{eqnarray}
where
\begin{equation}
\delta_Q^{\rm th,lin}(\tau) = \frac{1}{2 \pi^2} \int dk \ k^2 W(kX) \tilde\delta^{\rm lin}_Q(k,\tau)
\end{equation}
is the linearly evolved dark energy density contrast averaged over the top hat volume.

\section{Numerical results\label{sec:III}}

In this section we proceed to solve numerically the evolution
equations for the spherical collapse model presented earlier in
section~\ref{sec:II}.%
\footnote{The numerical code is written in C++ and employs GNU
scientific libraries for solving the evolution equations and for
interpolating the integral in equation~(\ref{eq:deltaqreal}).  A
verification of the convergence of equation~(\ref{eq:deltaqreal})
against the number and spacing of $k$-bins has been conducted.}
We assume a flat
spatial geometry for the universe so that the dark energy fraction
today is related to the dark matter fraction by $\Omega_Q = 1 -
\Omega_m$.  For the choice of the parameter $\Omega_m$  and the present Hubble
rate $H_0$, we use the WMAP 7-year best-fit values~\cite{wmap7}.  We consider only those
cases with constant $w$ and $c_s$, although our formulation is applicable also
to scenarios with time-dependent dark energy parameters.

We begin the evolution at a dimensionless time
coordinate of
\begin{equation}
  t_iH_0=2\times 10^{-6},
\end{equation}
corresponding to an initial scale factor of
\begin{equation}
a_i=a_0 \left(\frac{3t_iH_0\sqrt{\Omega_m}}{2}\right)^{2/3},
\end{equation}
if we assume $t_i$ to lie well within the matter domination epoch.
Taking an initial matter overdensity of %
$\delta^{\rm th}_{m,i}$
the initial value of the top hat radius can then be obtained
directly from equation~(\ref{eq:sphericalm}) given some mass $M$.
This mass, which we dub the ``halo mass'', is also the mass of dark matter
contained in the final collapsed object.%
\footnote{Note that the mass $M$ here refers to the mass of the dark matter component only,
although as suggested in~\cite{Creminelli:2009mu}, the true mass of the bound object should in
principle include  the contribution from the clustered dark energy component as well.}
Unless otherwise stated, the initial matter density contrast
is taken to be
\begin{equation}
\delta^{\rm th}_{m,i}=3\times 10^{-4}.
\end{equation}
We have chosen the above values for the initial time and matter density
contrast so that the collapse occurs at a time when the dark energy
component constitutes a significant part of the universe's energy
budget.

Since the top hat evolution
equation is a second order differential equation, we must also
specify the time derivative of $R$. This can be constructed by
differentiating equation~(\ref{eq:deltamth}) with respect to time.
Because the initial matter density contrast is much less than unity,
we can approximate $d\delta_m^{\rm th}/dt \simeq d\delta_m^{\rm
th,lin}/dt \simeq H \delta^{\rm th,lin}_m$ using linear perturbation
theory, and  thus,
\begin{equation}
\label{eq:drdt}
\left. \frac{1}{R}\frac{dR}{dt} \right|_{t_i}  \simeq \frac{2}{3t_i}\left(1-\frac{1}{3}\delta^{\rm th}_{m,i}\right).
\end{equation}
Finally the initial conditions for the dark energy evolution is
given in section~\ref{sec:initial}, particularly by the
interpolation formula~(\ref{eq:interpolate}).

\subsection{The collapse}

\begin{figure}[t]
\centering
\includegraphics[width=0.5\linewidth]{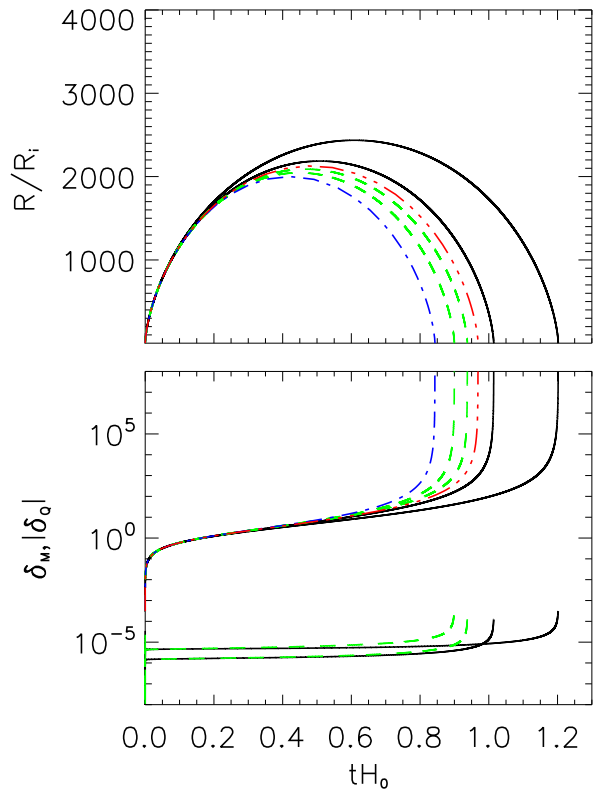}
\caption{{\it Top}: Spherical top hat radius $R$ normalised to the
initial radius $R_i$ as a function of the dimensionless time
coordinate $tH_0$.  The reference EdS and $\Lambda$CDM models are
represented by the blue dot-dash and the red dot-dot-dot-dash
lines respectively.  The two black solid lines denote, from right to
left, the cases of $w=-0.7,-0.9$, while the two green dashed lines
denote, from right to left, the cases of $w=-1.1,-1.3$. For
these four cases, we have chosen the dark energy sound speed to be
$c_s^2=10^{-1}$, and a halo mass of $M=10^{14}\Msun$. {\it Bottom}:
The corresponding top hat matter and dark energy density contrasts.
For the cases of $w=-1.1,-1.3$, the dark energy density
contrasts are negative, i.e., they are underdensities, and
the  $\delta_Q$ values presented in this plot are absolute values.
Note that in
the reference EdS and $\Lambda$CDM models, there is no dark energy
clustering. For all cases the initial matter overdensity has been
chosen to be $\delta^{\rm th}_{m,i}=3\times
10^{-4}$.\label{fig:sph3}}
\end{figure}

\begin{figure}[t]
\centering
\includegraphics[width=0.9\linewidth]{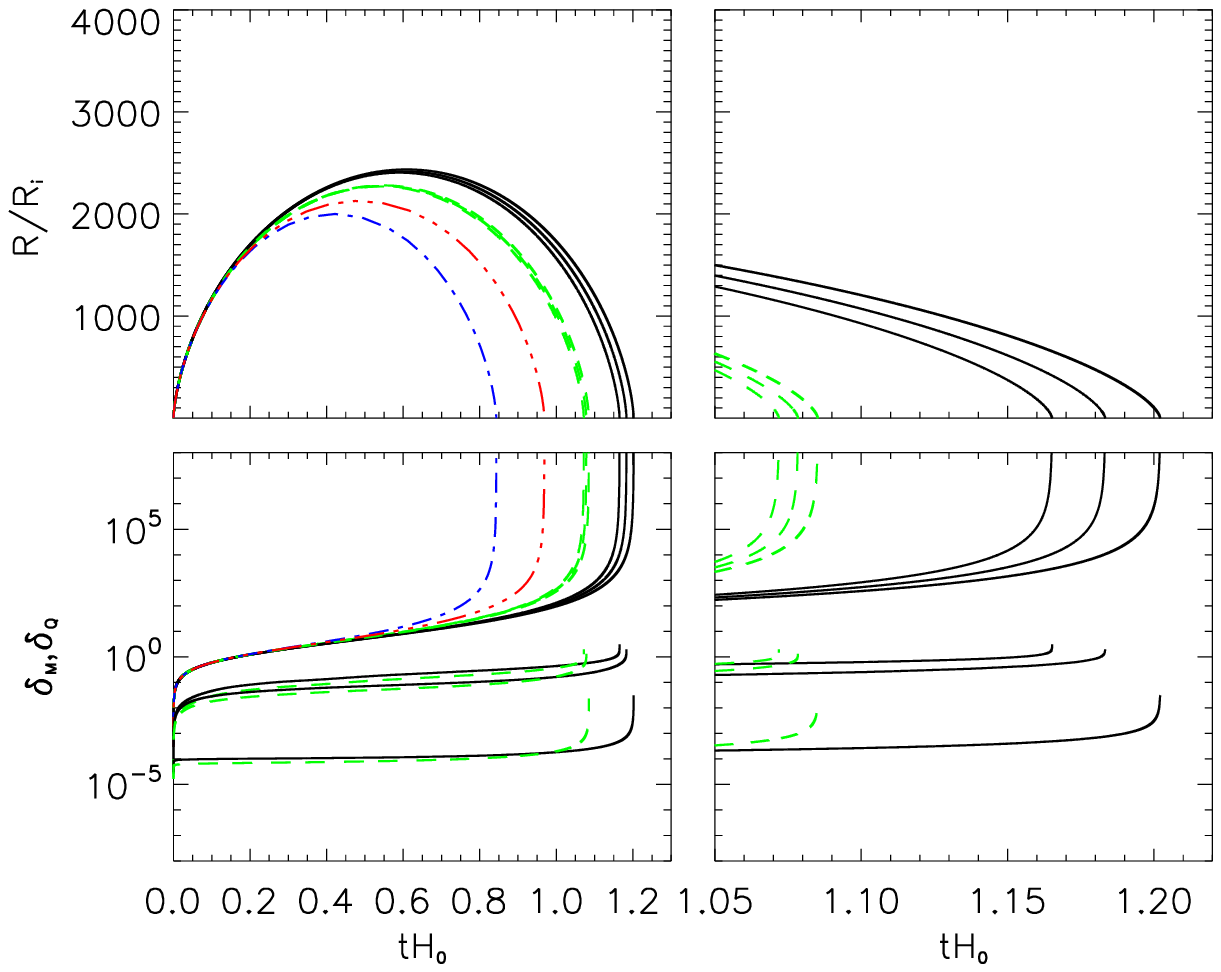}
\caption{Same as figure~\ref{fig:sph3}, but for a fixed halo mass
$M=10^{16}\Msun$. The black solid lines indicate $w=-0.7$ and, from
right to left, a dark energy sound speed of
$c_s^2=10^{-1},10^{-4},10^{-6}$. The green dashed lines denote
$w=-0.8$ for the same set of dark energy sound speeds. The reference
EdS and $\Lambda$CDM models are again represented by, respectively,
the blue dot-dash and the red dot-dot-dot-dash lines. The right
panels show the same results, but zoomed in on the time interval
$tH_0=1.05 \to 1.25$.\label{fig:sph1}}
\end{figure}

\begin{figure}[t]
\centering
\includegraphics[width=0.9\linewidth]{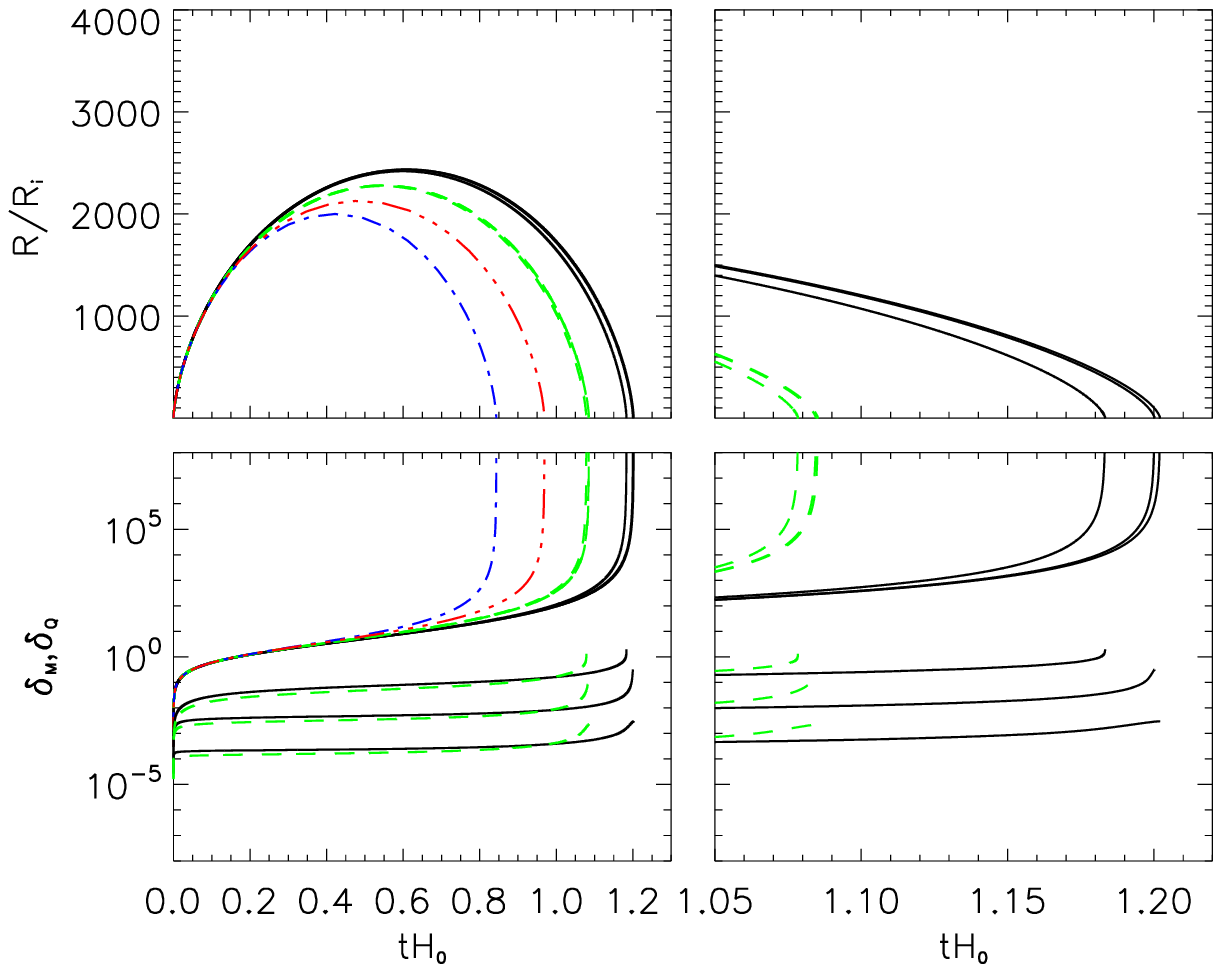}
\caption{Same as figure~\ref{fig:sph1}, but for a fixed sound speed
$c_s^2=10^{-4}$. The black solid lines denote $w=-0.7$ and, from
right to left, halo masses of $M=10^{12},10^{14},10^{16}M_\odot$. The green dashed lines
denote $w=-0.8$ for the same set of halo masses.
 The
blue dot-dash and the red dot-dot-dot-dash lines represent the
reference EdS and $\Lambda$CDM models respectively. The right panels
show the same results in the time interval between $tH_0=1.05$ and
$tH_0=1.25$.\label{fig:sph4}}
\end{figure}

\begin{figure}[t]
\centering
\includegraphics[width=0.9\linewidth]{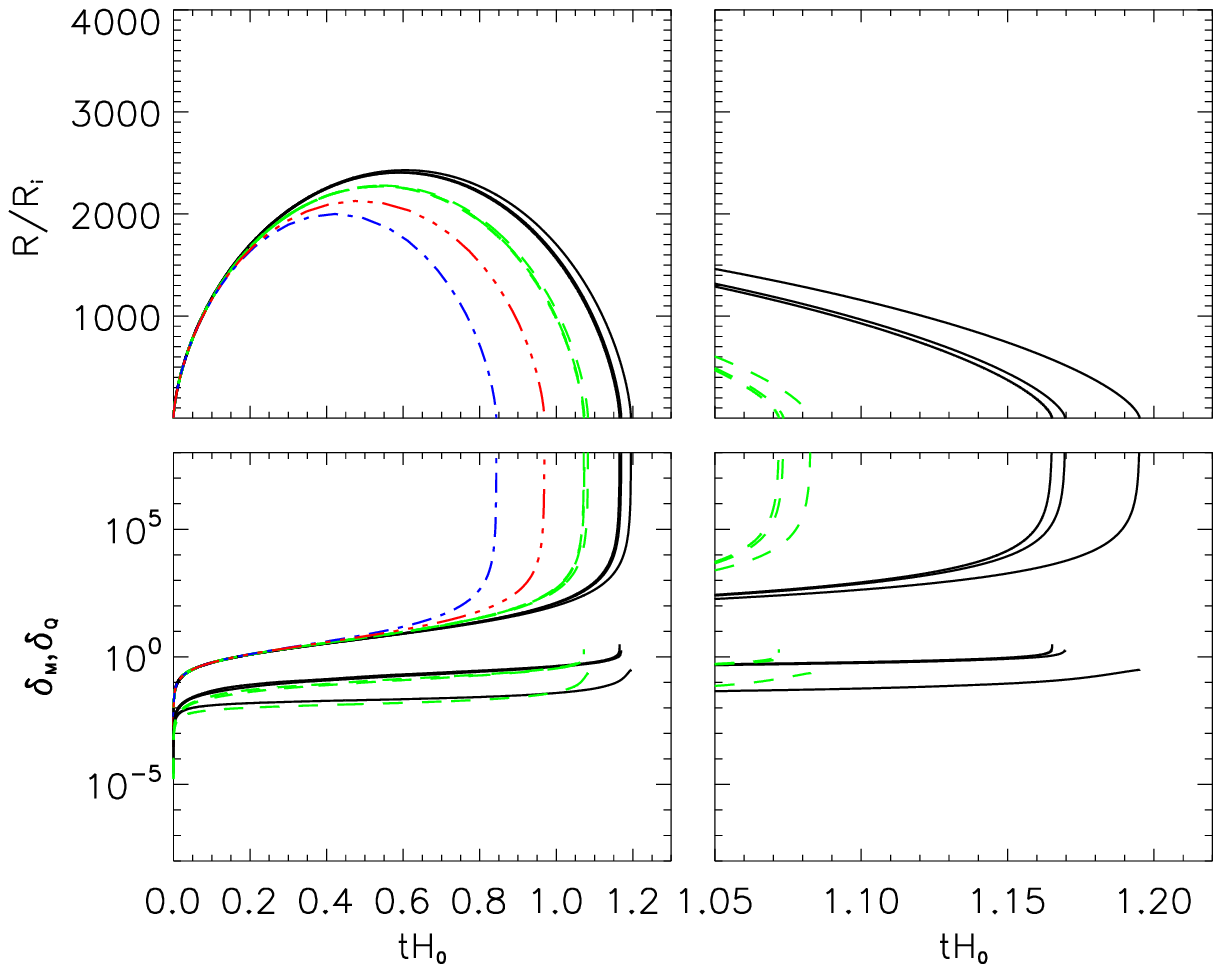}
\caption{Same as figure~\ref{fig:sph4}, but with the sound speed
fixed at $c_s^2 = 10^{-6}$.\label{fig:sph5}}
\end{figure}

Figures~\ref{fig:sph3} to~\ref{fig:sph5} shows the physical top hat radius
normalised to the initial radius as a function of the dimensionless
time coordinate $tH_0$ for several choices of $w$, $c_s$ and halo
masses $M$.
The corresponding matter
overdensity~(\ref{eq:deltamth}) and the dark energy
overdensity~(\ref{eq:deltaqreal}) are also shown in juxtaposition.

In figure~\ref{fig:sph3} we present the results for a dark energy
component with $c_s^2=10^{-1}$ and a halo mass of $M=10^{14}\Msun$
for various equation of state parameters. These choices of $c_s$ and
$M$ satisfy $M \ll M_J$ according to equation~(\ref{eq:jeansmass}),
and ensure that we are in the non-clustering regime.  Since dark
energy clustering is minimal ($< 10^{-5}$ relative to the dark
matter density contrast), only the equation of state parameter $w$,
i.e., the homogeneous part of the dark energy fluid, plays a role in
the dynamics of the spherical collapse.   Indeed, as we see in the
figure, the less negative the equation of state parameter, the later
the collapse.  The reason is that  for the same $\Omega_Q$, the less
negative $w$ is, the earlier the dark energy comes to dominate the
energy content of the universe, thereby inhibiting the growth of the
top hat overdensity through Hubble expansion from an earlier stage.
Comparing the $\Lambda$CDM case and a model with $w=-0.7$, the
collapse time is delayed by some 20\%. For equation of state
parameters more negative than $-1$, the opposite trend is seen; in
the case of $w=-1.3$ the collapse occurs some 10\% faster than in
the $\Lambda$CDM limit.

In figure~\ref{fig:sph1} we fix $M=10^{16}\Msun$, but vary the dark
energy sound speed and equation of state. We choose $w = -0.7$ and
$w = -0.8$, since, based on results from linear theory (see
section~\ref{sec:linear}), dark energy clustering is most enhanced
by a deviation of $w$ from $-1$ in the positive direction. Although
the effect of dark energy clustering on the spherical collapse is
quite small, the trend is clear: the smaller the sound speed, the
faster the collapse. This is to be expected, since the smaller the
sound speed is, the more efficiently the dark energy component
clusters, and this clustering in turn contributes to sourcing the
collapse of the top hat on the r.h.s.\ of
equation~(\ref{eq:xdotdot}).  Note that although the dark energy
component exhibits some degree of clustering in these cases, the
density contrast for almost the entire collapse history is quite
small until the last moments when $R \to 0$. This indicates that our
quasi-nonlinear approach---in which the dark energy component is
evolved with linearised equations---is valid for the model
parameters adopted in this figure.

Figure~\ref{fig:sph4} shows the cases of a fixed sound speed
$c_s^2=10^{-4}$, equation of state parameters $w=-0.7,-0.8$, and
three different halo masses. Figure~\ref{fig:sph5} is similar, but
with the sound speed  fixed at $c_s^2=10^{-6}$. For these sound
speeds, the corresponding Jeans masses $M_J$ are $4 \times
10^{17}M_\odot$ and $4 \times 10^{14} M_\odot$ respectively. In both
figures, we see that the larger the halo mass, the faster the
collapse. This can be understood from the non-clustering
solution~(\ref{eq:nonclustering}) (valid here since (almost) all
halo masses considered are less than the Jeans mass).  Since the
dominant Fourier mode is that corresponding to the comoving top hat
radius $X$ which is itself associated with the halo mass $M$, a
reasonable generalisation of the non-clustering
solution~(\ref{eq:nonclustering}) for the dark energy component in
terms of the halo mass would be
\begin{equation}
\delta^{\rm th}_Q(\tau) \sim \frac{3}{2} (1+w) \Omega_m(\tau) \left( \frac{M}{M_J}  \right)^{2/3} \delta_m^{\rm th} (\tau).
\end{equation}
The expression clearly shows that for a given sound speed, the
absolute value of the dark energy density contrast increases with
halo mass $M$. The enhanced dark energy density contrast in turn
hastens the collapse of the dark matter top hat.

Note that for the case of $M=10^{16} M_\odot$ and $c_s^2 = 10
^{-6}$, the dark energy density contrast is of order 0.1 for much of
the collapse history.  This suggests that our approximation scheme
for the dark energy evolution is approaching its limits of validity;
the approximation breaks down for larger masses.  Thus the rule of
thumb regarding the quasi-nonlinear approach appears to be that it
can be safely used for halo masses up to roughly the Jeans mass
$M_J$, but not beyond.%
\footnote{Obviously, we came to this
conclusion based on the rather extreme cases of $w=-0.7$ and
$w=-0.8$.  In general, however, we expect the validity of the
quasi-nonlinear approximation scheme to be dependent also on the
choice of $w$, where the less $w$ deviates from $-1$, the higher the
halo mass for which the approximation scheme remains applicable.}

\subsection{Linear threshold density\label{sec:deltalin}}

\begin{figure}[t]
\centering
\includegraphics[width=0.82\linewidth]{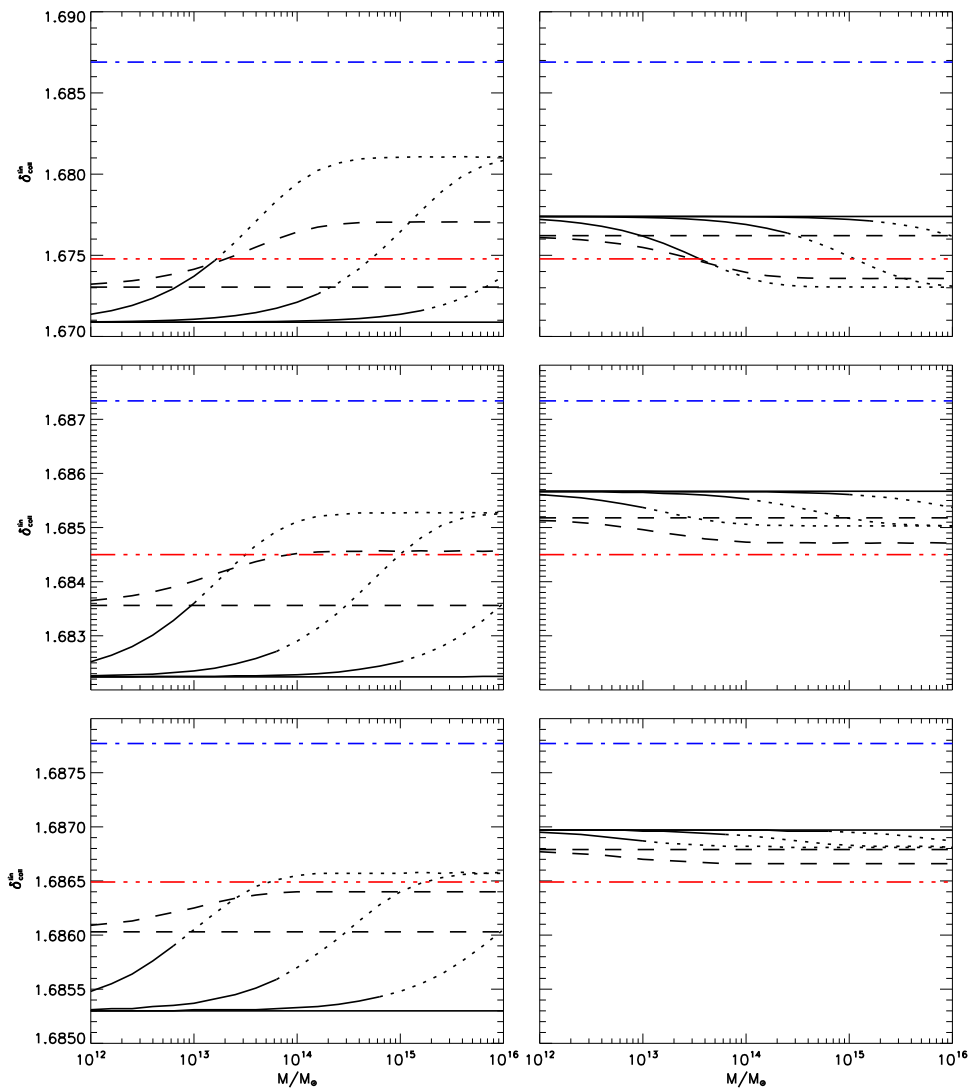}
\caption{Linear threshold density $\delta_{\rm coll}^{\rm lin}$ as a
function of the halo mass $M$ at various collapse redshifts.
{\it Top left}:  Collapse redshift of $z=0$.
Solid lines denote
models with $w=-0.8$ and, from top to bottom, $c_s^2=10^{-6},
10^{-5},10^{-4},10^{-2}$. Dashed lines denote  models with $w=-0.9$
and $c_s^2=10^{-6}, 10^{-2}$.   For those cases in which $\delta_Q^{\rm th}>1$ during the
collapse, we indicate the results with dotted lines.
The blue dot-dash and the red dot-dot-dot-dash lines
represent the reference EdS and $\Lambda$CDM models respectively.
{\it Top right}:  Same as top left, but with the solid
lines denoting models with $w=-1.2$ and, from top to bottom,
$c_s^2=10^{-2}, 10^{-4},10^{-5},10^{-6}$.  Dashed lines represent
models with $w=-1.1$ and $c_s^2=10^{-2},10^{-6}$.
{\it Middle}: Same as top panel, but for a collapse redshift of
 $z=1$.
{\it Bottom}: Same as top panel, but for a collapse redshift of
 $z=2$. \label{fig:delta}}
\end{figure}

Next, we compute the linear critical density contrast $\delta^{\rm
lin}_{\rm coll}$ required for use with such semi-analytic theories
as the Press--Schechter formalism and the excursion set theory. This
can be achieved by solving simultaneously {\it both}  the nonlinear
and the linear equations of motion for the spherical collapse, and
formally identifying  $\delta^{\rm lin}_{\rm coll}$ as the linearly
evolved matter density contrast at the instant the top hat radius
vanishes. Figure~\ref{fig:delta} shows $\delta^{\rm lin}_{\rm coll}$
as a function of the halo mass for various combinations of $w$ and
$c_s$. The initial matter overdensities are chosen such that  all
halos collapse at $z=0$ (top panel), $z=1$ (middle), and $z=2$
(bottom).

Clearly, for the reference cases of an EdS and a $\Lambda$CDM
universe, the linear threshold density is independent of the halo
mass.  However, once a finite dark energy sound speed is introduced
into the picture, $\delta^{\rm lin}_{\rm coll}$ becomes
mass-dependent, with $\delta^{\rm lin}_{\rm coll}(M)$ a
monotonically increasing function of $M$ when $w>-1$ and a
monotonically decreasing function of $M$ when $w<-1$. The
$M$-dependence is however quite weak for those cases with $w$ close
to $-1$, since dark energy clustering is generally suppressed by a
factor $(1+w)$.

The most interesting case presented here is that for $w=-0.8$,
especially for $c_s^2 = 10^{-5}$ and $c_s^2 = 10^{-6}$
(corresponding Jeans masses: $1.3 \times 10^{16}M_\odot$ and $4 \times
10^{14} M_\odot$). Here, we see that  at   $M \ll M_J$, $\delta^{\rm
lin}_{\rm coll}$ is at its lowest value and is essentially
independent of $M$, indicating that we are in the fully
non-clustering regime.  As we move to higher values of $M$, we
encounter a transition region where  $\delta^{\rm lin}_{\rm coll}$
rises with $M$.  Once $M \gg M_J$, however,  $\delta^{\rm lin}_{\rm
coll}$ reaches a plateau, where clustering is most efficient and
$\delta^{\rm lin}_{\rm coll}$ is again independent of $M$.
Interestingly, a similar pattern can also be seen in the $w<-1$ cases,
where the dark energy density contrasts are negative, corresponding
to dark energy underdensities, which have a
negative effect on the clustering of matter. This negative effect is strongest
for masses larger than the Jeans mass.

The dependence of $\delta_{\rm coll}^{\rm lin}$ on the collapse redshift is also
quite clear in figure~\ref{fig:delta}. The later the collapse, the larger the
difference between the values of  $\delta_{\rm coll}^{\rm lin}$ for the clustering
and the non-clustering solutions.  The reason is simply that
the contribution of dark energy to the total energy budget in the universe increases with time.
This means the effect of dark energy clustering also becomes more important at lower redshifts.

Finally, we caution the reader again that our quasi-nonlinear
approximation breaks down if the dark energy density contrast inside
the top hat becomes too large, especially when the halo mass
approaches or exceeds the Jeans mass associated with a chosen sound
speed.  As a rule of thumb, we take the condition of breakdown to be
$\delta_Q^{\rm th} \geq 1$ at any time during the collapse process.
To alert the reader to those cases where the breakdown condition is
met, we indicate the resulting linear threshold densities
$\delta^{\rm lin}_{\rm coll}$ in figure~\ref{fig:delta} with dotted
lines; in these cases, the exact values of $\delta^{\rm lin}_{\rm
coll}$ are unreliable. Clearly,  for equation of state parameters
that deviate significantly from $w=-1$ (e.g., $w=-0.8,-1.2$), our
approximate approach breaks down already at $M < M_J$. For
$w=-0.9,-1.1$, the approach appears to remain valid for a larger
mass range.

The breakdown of our approximation in the clustering limit also
explains why we do not recover the exact results of
reference~\cite{Creminelli:2009mu} for  $c_s^2 =0$ and
$w=-0.7,-1.3$. In fact, in the clustering limit,  our approximation
appears to overestimate the effect of dark energy clustering on the
linear threshold density; had we included all nonlinear effects,
nonlinear dark energy clustering would feed back on the matter
clustering more effectively, thereby leading to an earlier collapse
for $w>-1$ (and a later collapse for $w<-1$). An earlier collapse
means that the linearly evolved matter density contrast would reach
a lower value at the time of collapse, so that the real $\delta_{\rm
coll}^{\rm lin}$ in the clustering limit would be lower than our
estimate in figure~\ref{fig:delta}.

\subsection{Virialisation}

In reality the collapse of an overdense region will never take place
in the way described above, since density fluctuations inside the
region will moderate the infall, and the system reaches virial
equilibrium before the matter density  can ever become infinite. For
a single component system such as the case of a dark
matter-dominated EdS universe, the process of virialisation and the
radius at  which virial equilibrium is attained can be obtained
directly from the spherical collapse physics by assuming energy
conservation between the time of turnaround---defined as the
moment at which the top hat radius begins to shrink---and the time
at which virial equilibrium is established. The result turns out to
be rather simple,
\begin{equation}
R_{\rm vir}=\frac{1}{2}R_{\rm turnaround},
\label{eq:radvir}
\end{equation}
i.e.,  virialisation is complete by the time the top hat radius
decreases to half its value at turnaround.

For more complicated systems, such as the two fluid system
considered in this paper, the conditions of energy conservation need
to be modified. The time of virialisation can still be taken to be
the instant at which the virial theorem is satisfied.   However, the
problem is complicated by the fact that we do not know how or if the
dark energy fluid takes part in the virialisation process (see,
e.g., \cite{Maor:2005hq,Wang:2005ad}). Therefore, for simplicity, we
adopt equation~(\ref{eq:radvir}) as the virialisation condition for
this paper, and define the virial overdensity as
 \begin{equation}
 \label{eq:virdensity}
 \Delta_{\rm vir} \equiv \frac{\rho_m^{\rm th}(\tau_{\rm vir})}{\bar{\rho}_m(\tau_{\rm vir})},
  \end{equation}
where $\tau_{\rm vir}$ is the time at which $R=R_{\rm vir}$.
We expect the quantitative results to be somewhat
sensitive to our choice of the virialisation condition, but the
qualitative features should be unaffected.

\begin{figure}[t]
\centering
 \includegraphics[width=0.8\linewidth]{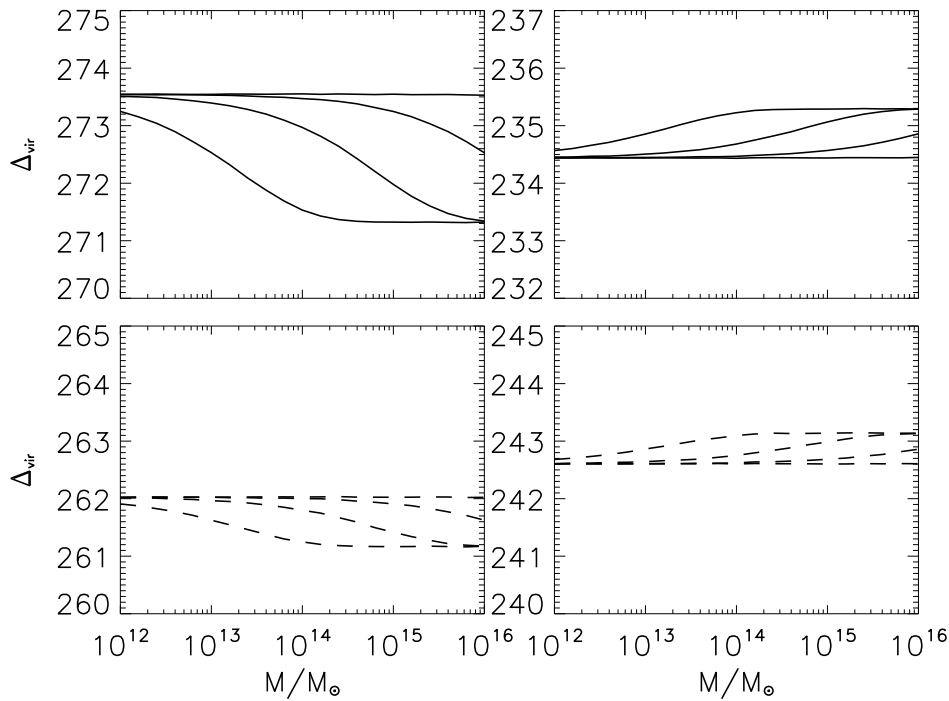}
 \caption{Virial overdensity $\Delta_{\rm vir}$ as a function of
 the halo mass $M$ for different dark energy equation of state parameters
 and sound speeds.  In all cases $\delta^{\rm
 th}_{m,i}$ has been chosen such that the collapse occurs at $z=0$.
  {\it Top left}: $w=-0.8$ and, from top to bottom, $c_s^2=10^{-2},
 10^{-4},10^{-5},10^{-6}$.
 {\it  Top right}:  $w=-1.2$ and $c_s^2=10^{-6}, 10^{-5},10^{-4},10^{-2}$.
  {\it Bottom left}:  $w=-0.9$ and $c_s^2=10^{-2}, 10^{-6}$.
  {\it Bottom right}:  $w=-1.1$ and
 $c_s^2=10^{-6}, 10^{-2}$.
 \label{fig:virial}}
\end{figure}

Figure~\ref{fig:virial} shows $\Delta_{\rm vir}$ as a function of
the halo mass $M$ for various combinations of $w$ and $c_s^2$. The
initial top hat matter density contrast $\delta^{\rm th}_{m, i}$ has
been fixed so that all halos collapse at $z=0$.   As a reference point,
for halos that collapse today,  $\Delta_{\rm vir}=147$ and $\Delta_{\rm vir}=252$
for an EdS and a $\Lambda$CDM universe respectively (we do not
plot these in figure~\ref{fig:virial} because they fall out of the plotting range).

Similar to the linear threshold density $\delta^{\rm lin}_{\rm
coll}$, $\Delta_{\rm vir}$ is first and foremost dependent on the
choice of $w$.  Introducing a finite dark energy sound speed into
the picture induces for $\Delta_{\rm vir}$ a dependence on the halo
mass $M$. However,  while  $\delta^{\rm lin}_{\rm coll}$ increases
with $M$ for cosmologies with $w>-1$, $\Delta_{\rm vir}$ decreases
with it. The opposite trend is seen for cosmologies with  $w<-1$. As
with $\delta^{\rm lin}_{\rm coll}$, again we see an asymptotic
non-clustering value for $\Delta_{\rm vir}$ at $M \ll M_J$, a
transition region at $M \sim M_J$ where $\Delta_{\rm vir}$ varies
strongly with $M$, and a second asymptotic region in the  $M \gg
M_J$ clustering limit.

The results in figure~\ref{fig:virial} are for a collapse redshift
of $z=0$. For halos that collapse earlier, the dependence of
$\Delta_{\rm vir}$ on the halo mass is qualitatively similar to the
$z=0$ case, but the difference in $\Delta_{\rm vir}$ between  the
clustering and the non-clustering limits is smaller. This trend is
reminiscent of the results in figure~\ref{fig:delta} for the linear
threshold density $\delta_{\rm coll}^{\rm th}$ for different
collapse redshifts.

Finally, note that  some authors define the virial overdensity as
the top hat density at the time of virialisation $\tau_{\rm vir}$,
but normalised to the background density evaluated at the {\it
collapse time} $\tau_{\rm coll}$.  This means that instead of, e.g.,
$\Delta_{\rm vir}=147$ for the EdS model according to our
definition~(\ref{eq:virdensity}), one finds a higher value of $179$
simply because between $\tau_{\rm vir}$ and $\tau_{\rm coll}$ the
background density has become smaller due to the Hubble expansion.
Since the dark energy sound speed does not play a role in the
background expansion, the effect of these differing definitions is
only to induce a shift in the normalisation of $\Delta_{\rm vir}$
for a given set of $w$ and $\Omega_Q$.  The $M$-dependence of
$\Delta_{\rm vir}$ is unaffected.

\section{Discussions and conclusions\label{sec:IV}}

While a dark energy fluid with a negative equation of state
parameter appears to describe the apparent accelerated expansion of
our universe with reasonable success, the precise nature and
properties of this dark energy remain undetermined. In this paper,
we have addressed some aspects of the dark energy's role in  cosmic
structure formation. Specifically, we have considered a generic dark energy
fluid parameterised by a constant equation of state parameter $w$
and sound speed $c_s$, and determined their impact on the formation
of  gravitationally bound objects.

Our main tool is the spherical collapse model,  incorporating a
nonrelativistic dark matter component and a generic dark energy
fluid described above.  Such a model has been investigated by other
authors previously in the limit where the dark energy is  (i)
non-clustering, i.e., $c_s \to \infty$, or (ii)  comoving with the dark
matter, i.e., $c_s \to 0$.   In this work, we have generalised the
spherical collapse model to describe those intermediate cases
characterised by an arbitrary $c_s$.   Although we have focussed on  scenarios
with constant $w$ and $c_s$,  our approach is equally applicable to
dark energy fluids described by time-dependent  parameters.
 Along the
way, we have also provided a detailed description of dark energy evolution
in the linear regime---again for arbitrary $w$ and $c_s$, and
identified some salient features of dark energy clustering.

The dark energy component changes the evolution of the spherical
collapse through its effect on the overall expansion of the universe
as well as through its own clustering abilities.   In the
non-clustering limit, only the equation of state parameter $w$ plays
a role in the dynamics of the spherical collapse by altering the
rate of the Hubble expansion.  When the dark energy is able to
cluster, however, then the dark energy density contrast also sources
the evolution of the dark matter density perturbations.  The amount
of dark energy clustering is determined by both $w$ and $c_s$.

In addition, since the introduction of a dark energy sound speed
$c_s$ necessarily brings into the picture a ``Jeans scale'' and hence
a ``Jeans mass'' $M_J$, we find that the same sound speed can
influence the spherical collapse dynamics in different ways
depending on the mass of the collapsed object or the ``halo mass''
$M$. This mass dependence is especially manifest when we compute the
virial overdensity $\Delta_{\rm vir}$ and the linearly extrapolated
threshold density $\delta_{\rm coll}^{\rm lin}$.  In both cases, we
find two asymptotic regions corresponding to the non-clustering $M
\ll M_J$ limit and the clustering $M \gg M_J$ limit, where the value
of $\Delta_{\rm vir}$ or $\delta_{\rm coll}^{\rm lin}$ is
practically constant with respect to $M$.  In between these limits
is a transition region, in which $\Delta_{\rm vir}$  ($\delta_{\rm
coll}^{\rm lin}$) declines (grows) strongly with $M$.  Observing
this transition region will be tell-tale sign that dark energy is
dynamic, and a great leap towards pinning down its clustering
properties.

One possible way to discern this halo mass-dependent dark energy clustering is to
measure the cluster mass function.  Already
it has been shown in, e.g., reference~\cite{Creminelli:2009mu}
that the difference between a clustering ($c_s=0$)
and a homogeneous ($c_s \to \infty$) dark energy fluid can be an order unity effect on
the expected number of clusters at the high mass tail of the  mass function.  It remains to be
seen how exactly an arbitrary dark energy sound speed would alter this conclusion, but some additional
scale-dependent effects are almost guaranteed to be present.

Finally, let us remind the reader again that in our spherical
collapse analysis we have simplified the equations of motion for the
dark energy component so that only terms linear in the dark energy
density contrast have been retained. This approach is strictly not
valid for those cases in which the dark energy density contrast
exceeds unity during the collapse process, and is especially prone
to breakdown for cosmologies with an equation of state parameter
that deviates significantly from $w=-1$. In these cases, a fully
nonlinear analysis is required in order to compute accurately such
quantities as $\Delta_{\rm vir}$  and $\delta_{\rm coll}^{\rm lin}$
for comparison with observations.  However,  the qualitative
features of dark energy clustering should not be affected by our
simplified approach.  A fully nonlinear analysis would require that
we solve the equations of motion for the dark energy component---two
$(1+1)$-dimensional partial differential equations---using either a
grid-based finite difference scheme or a Lagrangian method akin to a
smoothed particle hydrodynamics simulation.  Investigation is
already underway and we hope to report the results in a future
publication.

\ack

We thank Jacob Brandbyge, Jan Hamann and Cornelius Rampf for comments on the manuscript.
Y$^3$W thanks Guillermo Ballesteros for useful discussions. OEB acknowledges support from the Villum Foundation.

\appendix

\section{From general relativity to the pseudo-Newtonian approach\label{sec:rel}}

We demonstrate in this section that, at linear order, the pseudo-Newtonian approach adopted in this work is indeed consistent with the Newtonian limit of a general relativistic formulation.
We work in the conformal Newtonian gauge, in which the perturbed line element is given by~\cite{Ma:1995ey}
\begin{equation}
ds^2 = a^2(\tau) \{-[1+2 \psi(\tau,{\bm x})] d \tau^2+[1-2 \phi(\tau,{\bm x})] (dx^2+dy^2+dz^2)\},
\end{equation}
for a flat background spatial geometry.
Assuming zero anisotropic stress so that $\psi=\phi$, the equations of motion in Fourier space for the dark matter and dark energy components~\cite{Sapone:2009mb,Ballesteros:2010ks,Anselmi:2011ef}  are, 
respectively,
\begin{eqnarray}
\label{eq:ddmm}
&& \dot{\delta}_m + \theta_m - 3 \dot{\phi}=0, \nonumber \\
&& \dot{\theta}_m + {\cal H} \theta_m - k^2 \phi=0,
\end{eqnarray}
and
\begin{eqnarray}
\label{eq:ddee}
&&\dot{\delta}_Q + 3 (c_s^2 -w ) {\cal H} \delta_Q +(1+w) D  \theta_Q - 3 (1+w) \dot{\phi} =0, \nonumber \\
&&\dot{\theta}_Q + (1-3 c_s^2) {\cal H} \theta_Q - \frac{k^2 c_s^2}{1+w} \delta_Q - k^2 \phi=0,
\end{eqnarray}
where $D \equiv 1 + 9 (c_s^2 -w) {\cal H}^2/k^2$, and $\dot{D} = 18 (c_s^2 - w) \dot{\cal H} {\cal H}/k^2$.
Rearranging equations~(\ref{eq:ddmm}) and (\ref{eq:ddee}) respectively into second order differential equations, we find
\begin{equation}
\label{eq:dedm}
\ddot{\delta}_m + {\cal H} \dot{\delta}_m = 3 {\cal H} \dot{\phi} -k^2 \phi + 3 \ddot{\phi} ,
\end{equation}
and
\begin{eqnarray}
\label{eq:de2}
&& \ddot{\delta}_Q+ [(1-3w) {\cal H} - \dot{D}/D] \dot{\delta}_Q \nonumber \\
&& \hspace{10mm} + \{ D c_s^2 k^2 + 3(c_s^2 -w) [ \dot{\cal H} + (1-3 c_s^2) {\cal H}^2 - (\dot{D}/D) {\cal H}] \} \delta_Q \nonumber \\
&& \hspace{10mm} = (1+w) \{ 3 [(1-3 c_s^2) {\cal H} - \dot{D}/D] \dot{\phi} - D k^2 \phi + 3 \ddot{\phi} \},
\end{eqnarray}
assuming time-independent $w$ and $c_s^2$.

In order to take the subhorizon (i.e., $k\gg {\cal H}$) limit of equation~(\ref{eq:de2}), we note 
that $\dot{\cal H} \sim O({\cal H}^2)$.  Thus, we can replace all occurrences of $D$ with $D=1$ and
$\dot{D}/D$ with $\dot{D}/D=0$.  One last step concerns the coefficient of the $\delta_Q$ term on the LHS: we group all contributions proportional
to $c_s^2$ together to get
\begin{equation}
\label{eq:approximation3}
\left\{c_s^2 k^2 \left[ 1 + 3 \frac{{\cal H}^2}{k^2} \left(\frac{\dot{\cal H}}{{\cal H}^2} +1-3(c_s^2-w)\right) \right] - 3 w ({\cal H}^2+\dot{\cal H}) \right\} \delta_Q,
\end{equation}
and set  the  $O({\cal H}^2/k^2)$ terms in $[ \ldots]$ to zero.   Thus, we find
\begin{eqnarray}
\label{eq:de2a}
&& \ddot{\delta}_Q+ (1-3w) {\cal H} \dot{\delta}_Q + [c_s^2 k^2  -3w (\dot{\cal H} + {\cal H}^2)]\delta_Q \nonumber \\
&& \hspace{40mm} = (1+w) [3 (1-3 c_s^2) {\cal H} \dot{\phi} - k^2 \phi + 3 \ddot{\phi} ]
\end{eqnarray}
for the subhorizon limit of equation~(\ref{eq:de2}).

It remains to specify the metric perturbation $\phi$ in terms of the density and velocity perturbations via the Einstein equation.  
Using the expressions given in, e.g,~\cite{Ma:1995ey}, we obtain
\begin{eqnarray}
\label{eq:einstein}
&& k^2 \phi = - \frac{3}{2} {\cal H}^2 \sum_\alpha \Omega_\alpha \left[\delta_\alpha + 3 \frac{{\cal H}^2}{k^2} (1+w_\alpha) \frac{\theta_\alpha}{\cal H} \right], \nonumber \\
&&{\cal H} \dot{\phi} = \frac{3}{2} {\cal H}^2 \sum_\alpha \Omega_\alpha 
\frac{{\cal H}^2}{k^2} \left [\delta_\alpha +  (1+w_\alpha)\left(1+3 \frac{{\cal H}^2}{k^2}\right) \frac{\theta_\alpha}{\cal H} \right], \nonumber  \\
&& \ddot{\phi} = \frac{3}{2} {\cal H}^2 \sum_\alpha \Omega_\alpha \left\{\left[ c_\alpha^2+ \frac{{\cal H}^2}{k^2}  \left(\frac{2}{{\cal H}^2} \frac{\ddot{a}}{a} -4 \right) \right] \delta_\alpha
  \right. \nonumber \\
&& \hspace{20mm} + 3 (1+w_\alpha) \left. \frac{{\cal H}^2}{k^2} \left[ c_\alpha^2 -w_\alpha   -1+ \left(\frac{2}{{\cal H}^2} \frac{\ddot{a}}{a} -4\right)  \frac{{\cal H}^2}{k^2} \right] \frac{\theta_\alpha}{\cal H} \right\}. 
\end{eqnarray}
For the problem at hand,  the summation is performed over $\alpha =m,Q$, and the notation should be understood to mean $\Omega_\alpha = \Omega_\alpha(\tau)$, $w_m=0$, $w_Q=w$, $c_m^2=0$, and $c^2_Q = c^2_s$.  Combining these expressions to form the RHS of equation~(\ref{eq:dedm}), we find the leading order contribution in the subhorizon limit to be
\begin{equation}
\label{eq:aaa}
3 {\cal H} \dot{\phi} -k^2 \phi + 3 \ddot{\phi} 
 \simeq  \frac{3}{2} {\cal H}^2 \sum_\alpha \Omega_\alpha (1+3 c_\alpha^2) \delta_\alpha.
\end{equation}
Similarly, we find for the RHS of equation~(\ref{eq:de2}), 
\begin{equation}
\label{eq:bbb}
 3 (1-3c_s^2) {\cal H} \dot{\phi} -k^2 \phi + 3 \ddot{\phi}  \simeq  \frac{3}{2} {\cal H}^2 \sum_\alpha \Omega_\alpha (1+3 c_\alpha^2) \delta_\alpha,
\end{equation}
again to leading order in ${\cal H}/k$.  Note that RHS of equations~(\ref{eq:aaa}) and (\ref{eq:bbb}) are identical as a result of a ${\cal H}^2/k^2$-suppressed 
${\cal H} \dot{\phi}$ term relative to the $k^2 \phi$ and $\ddot{\phi}$ terms, as can be seen in equation~(\ref{eq:einstein}).  We have {\it not} assumed $c_s^2 \ll 1$ to arrive at this result.

Thus,  equations~(\ref{eq:dedm}) and (\ref{eq:de2}) now become, respectively,
\begin{eqnarray}
\label{eq:final}
&& \ddot{\delta}_m + {\cal H} \dot{\delta}_m =\frac{3}{2} {\cal H}^2 [ \Omega_m(\tau) \delta_m + \Omega_Q(\tau) (1+3 c_s^2) \delta_Q], \nonumber \\
&& \ddot{\delta}_Q+ (1-3w) {\cal H} \dot{\delta}_Q + [c_s^2 k^2  -3w (\dot{\cal H} + {\cal H}^2)]\delta_Q \nonumber \\
&&\hspace{30mm} = (1+w) \frac{3}{2} {\cal H}^2 [ \Omega_m(\tau) \delta_m + \Omega_Q(\tau) (1+3 c_s^2) \delta_Q].
\end{eqnarray}
These equations are consistent with the outcome of the pseudo-Newtonian approach (equations~(\ref{eq:lindeltam}) to (\ref{eq:dkx})).

\section{Validity of the ${\cal H} _s^2 \gg k^2 \gg {\cal H}^2$ regime\label{sec:validity}}

A number of approximations have been made in order to arrive at the final equation of motion~(\ref{eq:final}) for the dark energy component in the subhorizon limit.   We have demanded that 
$(1-3w) {\cal H} \gg  \dot{D}/D$ in equation~(\ref{eq:de2}), which limits the use of the approximate equation~(\ref{eq:final}) to 
${\cal H}^2/k^2 \ll 0.074 \to 0.15$,
depending on the sound speed assumed.  We have also assumed $D \approx 1$,  equivalent to imposing a limit of
${\cal H}^2/k^2 \ll 0.055 \to 0.1$ on the validity of our approximate equations.  In practice, we might want to choose a benchmark limit of
\begin{equation}
\label{eq:subhorizon}
{\rm max} \left(\frac{{\cal H}^2}{k^2} \right) = 0.01,
\end{equation}
in order to keep all superhorizon contributions at a truly subdominant level.

Similarly, if we wish to take the super-sound-horizon limit (i.e., the clustering limit, $k\ll {\cal H}_s$, where ${\cal H}_s \equiv {\cal H}/c_s$), then from equation~(\ref{eq:harmonic})
or (\ref{eq:final}), we see that 
\begin{equation}
\label{eq:supersound}
{\rm max}  \left(\frac{k^2 c_s^2}{{\cal H}^2} \right) = 0.1
\end{equation}
would work as a good benchmark limit to ensure the subdominance of sub-sound-horizon contributions.  The two benchmark limits~(\ref{eq:subhorizon}) and 
(\ref{eq:supersound}) 
together define the ${\cal H} _s^2 \gg k^2 \gg {\cal H}^2$ regime (or the ``clustering regime'' in this work). 

From here,  it is easy to see that the two limits~(\ref{eq:subhorizon}) and (\ref{eq:supersound})  combine to set 
a constraint on the  dark energy sound speed  of
\begin{equation}
c_s^2 \lwig 10^{-3}
\end{equation}
in the  ${\cal H} _s^2 \gg k^2 \gg {\cal H}^2$ regime.  If we were to exceed this constraint, we also run the risk of increasing the superhorizon (${\cal H}^2/k^2$) and/or sub-sound-horizon ($k^2 c_s^2/{\cal H}^2$) contributions to any approximate analytic solution to beyond the ``negligible'' level, in which case the ${\cal H} _s^2 \gg k^2 \gg {\cal H}^2$ regime becomes ill-defined.

\end{document}